\documentclass[twocolumn,           
               showpacs,            
               preprintnumbers,     
               aps,                 
               prd,                 
               a4paper,             
               superscriptaddress,  
               nofootinbib,         
               tightenlines,        
               floats,floatfix,     
               amsmath,amssymb
               ]{revtex4}

\usepackage{graphicx}
\usepackage{subfigure}
\usepackage{latexsym}
\usepackage{amsmath,amssymb}        
\usepackage[draft=false]{hyperref}

\begin{document}



\title{Bose-Einstein condensation of relativistic Scalar Field Dark
  Matter}
\author{L. Arturo Ure\~na-L\'opez}
\affiliation{Instituto de F\'isica de la Universidad de Guanajuato,
C.P.~37150, Le\'on, Guanajuato, M\'exico}  
\date{\today}
\pacs{98.80.-k, 03.75.Nt, 95.35.+d \hfill arXiv:}
\preprint{arXiv:}


\begin{abstract}
Standard thermodynamical results of ideal Bose gases are used to study
the possible formation of a cosmological Bose-Einstein condensate in
Scalar Field Dark Matter models; the main hypothesis is that the boson
particles were in thermal equilibrium in the early Universe. It is
then shown that the only relevant case needs the presence of both
particles and anti-particles, and that it corresponds to models in
which the bosonic particle is very light. Contrary to common wisdom,
the condensate should be a relativistic phenomenon. Some cosmological
implications are discussed in turn.
\end{abstract}

\maketitle


\section{Introduction}
Scalar Field Dark Matter (SFDM)
models\cite{Sin:1992bg,Sahni:1999qe,Matos:2000ng,Hu:2000ke,Arbey:2001qi},
in which the dark matter (DM) particle is a spin-$0$ boson, are
becoming a serious alternative to the Cold DM paradigm that invokes
the existence of weakly interacting massive particles
(WIMP's)\cite{Taoso:2007qk}.

The WIMP hypothesis has been widely studied in many proposed
extensions of the Standard Model of Particle Physics, and there are
high expectations for its detection in the near
future\cite{Bertone:2004pz}. Support for the WIMP
hypothesis comes from the comparison between high-resolution N-body
simulations and the large scale structure we observe in the
Universe. The resulting DM distribution seems to be in good agreement
with observations\cite{DelPopolo:2008mr}, from clusters of galaxies
down to the scales of the galaxies themselves.

Good agreement means that there are fine details for which the WIMP
scenario does not have a definite answer. Those details refer to the
internal DM density profile in galaxies, the low number of galaxy
satellites, etc., see for
instance\cite{Tasitsiomi:2002hi}. It is not clear if the
observed discrepancies are due to corrections to the WIMP hypothesis
that have not been taken into account yet, but the possibility exists
that the DM particle is something rather different to a WIMP.

It is here where SFDM models enter into play. The usual approximation
in cosmological studies so far is to describe SFDM particles by means
of a scalar field $\phi$ minimally coupled to gravity that is endowed
with a scalar field potential $V(\phi)$. The dynamics of any SFDM
model depends upon the particular form of the scalar potential, but it
is well known that a massive potential as simple as $V(\phi) = m^2
\phi^2 /2$, where $m$ is the mass parameter of the SFDM particles, is
enough to successfully describe DM in a cosmological setting up to
linear order in the formation of large scale
structure\cite{Hwang:1996xd,Matos:2000ss,Matos:2008ag}. The key point
here is that the scalar field oscillations around the minimum of the
aforementioned quadratic potential make the SFDM particles behaves as
cold DM candidates.

More difficult tasks are those of non-linear structure formation,
though some progress in that direction has been recently
reported\cite{Woo:2008nn}. Other studies show
that the scalar field $\phi$ can form gravitationally stable
structures and that some of them may explain the particular features
observed in the rotation curves of
galaxies\cite{Sin:1992bg,Arbey:2001qi,Matos:2007zza}. Nonetheless,
the full formation of cosmological structure still is an open question
in SFDM models.

The mass $m$ is a free parameter that has to be constrained by
cosmological observations. The values considered in the literature
cover an ample range of masses, from very heavy values of the order of
$10^{13} \, \rm{GeV}$ to values as light as $10^{-23} \, \rm
{eV}$\cite{Sin:1992bg,Sahni:1999qe,Matos:2000ss,Arbey:2001qi,Liddle:2006qz}. It
is usually agreed that very heavy SFDM particles would be
indistinguishable from WIMP's in many respects, and the reason is that
cosmological bosonic features show up only if the Compton length of
the scalar field $\lambda_C \equiv m^{-1}$ is of the order of galactic
dimensions. This is usually the case in galactic studies, in which
very light values of the scalar field mass are ubiquitous.

On the other hand, the existence of SFDM particles, because of its
bosonic nature, opens the possibility for the formation of a
cosmologically relevant Bose-Einstein condensate
(BEC)\cite{Sin:1992bg,Bernstein:1990kf,Parker:1991jg,Ferrer:2004xj,Matos:2008ag}. Actually,
the scalar field $\phi$ mentioned before is supposed to represent the
wave function that describes the evolution of a cosmological BEC, see
for instance\cite{Parker:1991jg}, and the scalar
potential $V(\phi)$ encloses, in general, all the possible
self-interaction terms that may be present among the SFDM
particles. Therefore, the scalar field description of DM and the
formation of a BEC are closely related, but their relation is yet to
be fully understood\cite{Ferrer:2004xj}.

In this paper, I study the conditions for the formation of a
cosmological SFDM BEC using known results of the thermodynamics of
ideal Bose gases. I shall show that the formation of a BEC is not a
generic process in the history of the Universe, but one that only
appears for a relativistic SFDM gas composed of very light massive
particles.

To set up the general framework, I describe first the general
assumptions that will be present throughout the calculations
below. First of all, the SFDM particles are supposed to have been in
thermal equilibrium with other matter components in the early
universe, at a time when they were relativistic. That means the
temperature $T_i$ of the early thermal bath was such that $m \ll T_i
\simeq 10^9 \, \rm{GeV}$, as this is the usual initial value assumed
for the temperature at the beginning of the so-called Hot Big Bang.

Moreover, the SFDM particles will also decouple from the thermal bath
while still relativistic. That is, the temperature at decoupling $T_D$
is much larger than the mass of the particles, $m \ll T_D <
T_i$. Because we expect the Universe to expand adiabatically, the
temperature of the SFDM gas will always be very similar to the
temperature of the Universe itself.

Another assumption is that SFDM particles may or may not become
non-relativistic at some point in the evolution of the Universe. This
will allow us to study a ample range of masses for the SFDM particles,
including mass values that are well below the present temperature of
the Universe, $m < T_0 \simeq 10^{-4} \, \rm{eV}$. 

We shall find that, for all practical purposes, the number density of
SFDM particles at any given time suffices to decide about the
formation of a BEC. Other thermodynamical quantities will be
needed at some point, but only to clarify some features about the
expansion of the Universe. All calculations below are given in units
for which $\hbar = c = k_B = 1$.

A brief description of the paper is as follows. In
Sec.~\ref{sec:stand-therm-hist} I revise the formation of a BEC using
standard formulae for Bose ideal gases; the same is done in
Sec.~\ref{sec:thermal-history-sfdm} but with formulae that include the
coexistence of particles and anti-particles. Finally,
Sec.~\ref{sec:conclusions-} is devoted to conclusions.

\section{Standard thermal history of
  SFDM \label{sec:stand-therm-hist}}
I start with standard textbook results\cite{Kolb:1990vq} to calculate
the relic abundance of relativistic SFDM particles that become
non-relativistic at late times; the latter is a necessary condition
for them to be \emph{cold} DM particles. The number density $n_\phi$
of an ideal Bose-Einstein gas of SFDM particles
is\cite{PhysRev.138.A1049}
\begin{equation}
  n_\phi (T,\mu) = \frac{1}{2 \pi^2} \int^\infty_0 \frac{k^2
    dk}{\exp{\left[\beta (E_k-\mu(T))\right]}-1} \,
  , \label{eq:nphi-0}
\end{equation}
where $E^2_k = k^2+m^2$, $\beta=1/T$ is the usual Boltzmann factor,
$T$ is the temperature of the thermal bath, and $\mu(T)$ is the
chemical potential. Relativistic means that $m \ll T$; in addition, we
must impose the condition $|\mu| < m$ in order to keep $n_\phi$
positive definite\cite{Haber:1981fg}.

After the decoupling of the SFDM particles, the form of the distribution
function remains that of a relativistic (massless) component, and
the temperature of the SFDM particles, that I will denote by $T_\phi$
henceforth, redshifts strictly as $T_\phi \propto a^{-1}$, where $a$
is the scale factor of the universe. For the rest of the species which
continue to be in thermal equilibrium, the temperature falls a bit
slower since entropy conservation implies that $T \propto g^{-3}_S(T)
a^{-1}$, where $g_S(T)$ represents the entropy degrees of freedom at
temperature $T$. The number density~(\ref{eq:nphi-0}) then acquires
the usual form of a relativistic bosonic component
\begin{equation}
n_\phi = \frac{\zeta(3)}{\pi^2} T^3_\phi \, , \label{eq:nphi-1}
\end{equation}
where $\zeta(x)$ is the Riemann zeta function of $x$. 

It is useful to define the SFDM mass per photon $\xi_\phi \equiv
\rho_\phi/n_\gamma$\cite{Tegmark:2005dy}, where $\rho_\phi$ denotes
the energy density of the SFDM. While in the relativistic regime, $m
\ll T_\phi$, $\xi_\phi$ evolves as
\begin{equation}
  \xi_\phi = \frac{\pi^4}{60 \zeta(3)} \frac{T^4_\phi}{T^3} \,
  . \label{eq:xiphi}
\end{equation}
Both temperatures $T_\phi$ and $T$ evolve approximately at the same
rate, and then we conclude that $\xi_\phi \sim a^{-1}$.

Later on in the history of the Universe, the SFDM particles become
\emph{non-relativistic} once $T_\phi \simeq m$. Their energy is $E_k
\simeq m$, and then the energy density is simply given by $\rho_\phi
\simeq m \, n_\phi$. Therefore, the SFDM mass per photon is now given
by
\begin{equation}
  \xi_\phi = \frac{m}{2} \frac{T^3_\phi}{T^3} = \frac{m}{2}
  \frac{g_S(T_D)}{g_S(T)} \, . \label{eq:xiphi-1}
\end{equation}
The last equality in the above equation arises from the fact that the
SFDM particles were relativistic at the time they decoupled from the
thermal bath.
 
If we take the present measured value $\xi_{\phi,0} = 3.3 \times
10^{-28} \, m_{\rm{Pl}}$\cite{Tegmark:2005dy}, Eq.~(\ref{eq:xiphi-1})
imposes tight constraints on the mass value of any SFDM particle that
decoupled at a time when it was relativistic. Taking standard values
such as $g_S(T_D) \simeq 10^2$ and $g_S(T_0) \simeq
3.6$\cite{Kolb:1990vq}, where the
subscript '$0$' denotes present values, we find that the mass of the
SFDM particles should be $m \simeq 1 \, \rm{eV}$. This particle became
non-relativistic at a redshift of $z \simeq m / T_0 \sim 10^{4}$, just
on time to become a cold DM candidate.

The simple calculation above, which parallels those on generic
hot DM relics\cite{Kolb:1990vq},
readily shows that present data would prefer a SFDM candidate as a
massive as the neutrino, and that it would be quite difficult to
accommodate either heavier or lighter SFDM particles that were in
thermal equilibrium in the early universe, unless the entropy degrees
of freedom at the time of decoupling $g_S(T_D)$ take some unexpected
and extreme values. But this seems to be very unlikely.

It is now time to turn our attention to the non-standard textbook case
in which part of the SFDM particle budget is in the form of a BEC. As
we said before, because the SFDM were relativistic at decoupling, the
evolution of $n_\phi$ is given by Eq.~(\ref{eq:nphi-1}) at all later
times.

We then have to take a look at the relativistic conditions for the
formation of a BEC\cite{PhysRev.138.A1049}; in this case, the critical
temperature $T_c$ and the critical number density $n_{\phi,c}$ are
given, respectively, by
\begin{equation}
  T_c = \left[ \frac{\pi^2}{\zeta(3)} n_ \phi \right]^{1/3} \, , \quad
  n_{\phi,c} = \frac{\zeta(3)}{\pi^2} T^3_\phi \, . \label{eq:bec-nrel2}
\end{equation}
Hence, we can say that the formation of a BEC proceeds for
temperatures $T_\phi < T_c$, or for number densities such that
$n_{\phi,c} < n_\phi$. 

In other words, Eq.~(\ref{eq:bec-nrel2}) gives, for a given
temperature, the maximum number density that can be accounted by the
excited states; then, $n_{\phi,c} < n_\phi$ means that some particles
must necessarily reside in the (condensed) ground state.

A quick comparison of Eqs.~(\ref{eq:bec-nrel2}) with
Eq.~(\ref{eq:nphi-1}) shows that the excited states are just able to
accommodate all SFDM particles, and then no BEC is necessary.

\section{Thermal history of SFDM
  revisited \label{sec:thermal-history-sfdm}}
I now show that the formation of a relativistic BEC is possible if we
take into account the coexistence of particles and anti-particles in a
relativistic ideal Bose gas\cite{Haber:1981fg,Bernstein:1990kf} (see
also\cite{Kolb:1990vq}). To begin with, the equation that
replaces~(\ref{eq:nphi-0}) is
\begin{widetext}
  \begin{equation}
    q_\phi (T,\mu) = \frac{1}{2\pi^2} \int^\infty_0 k^2 dk \left[
      \frac{1}{\exp{[\beta (E_k - \mu(T))]}-1} - \frac{1}{\exp{[\beta
          (E_k + \mu(T))]}-1} \right] \, , \label{eq:nphi-rel1}
  \end{equation}
\end{widetext}
where we are explicitly assuming the presence of SFDM anti-particles
with a chemical potential $-\mu$. Following the standard notation, I
will use $q_\phi$ hereafter to denote the total \emph{charge} of the
SFDM.

The properties of the charge density~(\ref{eq:nphi-rel1}) have been
widely explored in the literature. For our purposes, it suffices to
know that at high temperatures, $m \ll T$, the charge density in
excited states is\cite{Haber:1981fg}
\begin{equation}
  q_\phi = \frac{\mu(T_\phi)}{3} T^2_\phi \, . \label{eq:nphi-rel2}
\end{equation}
The above formula is true only for the case $\mu (T_\phi) \leq m$, as
for larger values of the chemical potential the charge density is not
positive definite. We conclude that the maximum charge density allowed
by the excited states at a given relativistic temperature is $q_\phi
(T_\phi) = m T^2_\phi/3$. 

We have then given the needed description to define a critical
temperature $T_c$ and a critical charge density $q_{\phi,c}$ for the
formation of a BEC in an ultra-relativistic Bose gas of SFDM
particles. The new formulae that replace Eqs.~(\ref{eq:bec-nrel2}) are
\begin{equation}
  T_c = \left( \frac{3 q_\phi}{m} \right)^{1/2} \, , \quad q_{\phi,c}
  = \frac{m}{3} T^2_\phi \, . \label{eq:bec-rel1}
\end{equation}

On the other hand, cosmological expansion is adiabatic, and then we
expect the usual scaling of the SFDM charge density $q_\phi =
\eta_\phi T^3_\phi$ after decoupling, where $\eta_\phi$ is an appropriate
constant we determine below. The reason is that the entropy density of
an ultra-relativistic ideal Bose gas is
\begin{equation}
  s_\phi(T_\phi) = \frac{4\pi^2}{45} T^3_\phi \, . \label{eq:entropy-rel}
\end{equation}
If the evolution of SFDM particles proceeds separately at constant
entropy, $S_\phi = s_\phi a^3 = \rm{const}$, we find the usual
relationship $a \, T_\phi = \rm{const}$.

The condition $q_{\phi,c} < q_\phi$ for the formation of a BEC
translates into
\begin{equation}
  T_\phi > T_{\phi,c} \equiv \frac{m}{3\eta_\phi} \, . \label{eq:tphic}
\end{equation}
In the same way, if we solve the first of Eqs.~(\ref{eq:bec-rel1}) for
the critical temperature, we find
\begin{equation}
  T_c(T_\phi) = \left( \frac{T_\phi}{T_{\phi,c}} \right)^{1/2} \,
    T_\phi \, . \label{eq:tc-rel}
\end{equation}
As long as Eq.~(\ref{eq:tphic}) is satisfied, the above equation
confirms the standard wisdom that $T \ll T_c$ is necessary for the
formation of a BEC.\footnote{It is said in
  Refs.\cite{Bernstein:1990kf,Parker:1991jg} that the
  critical   temperature is $T_c=m/3\eta_\phi$. This statement is
  confusing, as the very definition of $T_c$ is
  Eq.~(\ref{eq:bec-rel1}) for a given value of the charge density
  $q_\phi$. Thus, my interpretation is that Eq.~(\ref{eq:tphic})
  determines the minimum temperature $T_{\phi,c}$ of the SFDM
  particles at which a cosmological BEC can form.}

As shown in\cite{Parker:1991jg}, the measurements on the
present contribution of the DM can help to determine the
$\eta_\phi$-constant that appears in Eq.~(\ref{eq:tphic}). We expect
the present charge density in SFDM particles to be of the order of
$q_{\phi,0} \simeq n_{b,0}$, where $n_{b,0} \simeq 10^{-10}
n_{\gamma,0}$ is the present baryon charge density. Finally, under the
assumption that the SFDM particles were once part of the thermal bath
in the early universe, a good estimation is
\begin{equation}
  \eta_\phi \simeq 10^{-10} \frac{\zeta(3)}{\pi^2}
  \frac{g_S(T_D)}{g_S(T_0)} \simeq 10^{-10} \,
  . \label{eq:etaphi}
\end{equation}

Once the BEC is formed in the early Universe at high temperatures, its
corresponding charge density $q_{BEC}$ is
\begin{equation}
  q_{BEC}(T_\phi) = n_{\phi} (T_\phi) - \frac{m}{3} T^2_\phi =
  \eta_\phi T^3_\phi \left( 1- \frac{T_{\phi,c}}{T_\phi} \right) \,
  . \label{eq:nbec-rel}
\end{equation}
The BEC charge density in the above equation is equivalent to other
results obtained
before\cite{Haber:1981fg}; to show the
equivalence one only needs to take the critical temperature $T_c$
defined in Eq.~(\ref{eq:tc-rel}) and substitute it in
Eq.~(\ref{eq:nbec-rel}).

\subsection{Non-relativistic SFDM}
Even if I have thoroughly supposed the existence of relativistic SFDM particles
in the early universe, I include here, for completeness, a short
description of the case in which SFDM particles are non-relativistic
at decoupling.

It is well known that the density of particles for a decoupled
non-relativistic species, for which $m \gg T_D$, is given
by\cite{Kolb:1990vq}
\begin{equation}
  n_\phi = e^{-\left( m - \mu_D \right)/T_D} \left( \frac{m T_\phi}{2\pi}
  \right)^{3/2} \, , \label{eq:nphi-nrel}
\end{equation}
where we notice an exponential suppression in which $\mu_D$ and $T_D$
are the values of the chemical potential and of the temperature at the
time of decoupling, respectively. It must be noticed that the
non-relativistic expression above is the same for the two relativistic
cases discussed previously\cite{Haber:1981fg}.

The non-relativistic conditions for the formation of a BEC
are\cite{PhysRev.138.A1049}
\begin{equation}
  T_c = \frac{2\pi}{m} \left[ \frac{n_\phi}{\zeta(3/2)} \right]^{2/3}
  \, , \quad  n_{\phi,c} = \zeta(3/2) \left( \frac{m \, T_\phi}{2\pi}
  \right)^{3/2} \, . \label{eq:tc-nrel}
\end{equation}
Because of the constraint $|\mu| \leq m$, a comparison of
Eqs.~(\ref{eq:nphi-nrel}) and~(\ref{eq:tc-nrel}) leads us to the
conclusion that a BEC does not appear for a non-relativistic species;
this is but the known result that we cannot change between the
uncondensed and condensed phases in an adiabatic
process\cite{Haber:1981fg}.

\section{Conclusions \label{sec:conclusions-}}
I have described above the main criteria that can lead to the
formation of a cosmological BEC within the SFDM hypothesis. The
different formulae presented must give a correct description of the
SFDM thermodynamics as long as there are no sudden critical changes in
the evolution of the Universe that may require a separate
study\cite{Kolb:1990vq}. At the same order of approximation, the
effects of the expansion of the Universe are taken into account in the
adiabatic behavior of the thermodynamical quantities.

First of all, it is remarkable that the assumption of thermal
equilibrium of the SFDM particles imposes tight constraints in the
formation of a cosmological BEC. In general, the condensed phase is
unreachable because of the adiabatic expansion of the universe.

The existence of a BEC is possible if the presence of anti-particles
is explicitly taken into account. The discussion leaded us to
Eq.~(\ref{eq:tphic}), a simple formula that gives the threshold
temperature $T_{\phi,c}$ for the formation of a BEC in terms of basic
physical quantities, the mass $m$ of the SFDM particles and their net
charge contribution, see $\eta_\phi$ in Eq.~(\ref{eq:etaphi}), to the
total matter contents of the universe.

On the other hand, Eq.~(\ref{eq:tphic}) points out that for massive
enough SFDM particles, the cosmological BEC dissapears at late times;
the reason lies in the fact that the SFDM gas is diluted by the
expansion of the Universe. Actually, we can easily estimate the mass
values for which a SFDM BEC exists up to the present time. By taking
$T_{\phi,c} = T_0$ in Eq.~(\ref{eq:tphic}), we find that a BEC always
exists for $m < 10^{-14} \, \rm{eV}$. It is exactly this latter case
in which a cosmological scalar field $\phi$ can be considered the
right description of SFDM.

For particles with $m > 10^{-14}\, \rm{eV}$, the SFDM gas is still
relativistic at the temperature at which the BEC disappears, $m \ll
T_{\phi,c}$, see Eq.~(\ref{eq:tphic}). After that, the total charge
density is entirely accounted by the the excited states, $q_\phi =
\eta_\phi T^3_\phi = \mu(T_\phi) T_\phi /3$, from which we find that
the functional form of the chemical potential is the expected
adiabatic one, $\mu(T_\phi) = 3 \eta_\phi
T_\phi$\cite{Bernstein:1990kf,Haber:1981fg}.

If $m > 1 \rm{eV}$, the SFDM gas should also become non-relativistic
before the present time. But because its number density continues to
be that of an ultra-relativistic species, no BEC will be further form
at low temperatures. A similar result arises for very large masses so
that the SFDM particles decoupled while non-relativistic; the
formation of a BEC is not allowed by the adiabatic expansion of the
universe\footnote{Of course, the formation of a BEC should be
  accomplished, in both cases, in the limit $T=0$ for which the
  excited states are all depopulated.}.

It is interesting that the existence of a BEC in a
cosmological setting requires both high temperatures and relativistic
particles. This is contrary to usual expectations according to which
the formation of a BEC is considered a low-temperature and
non-relativistic phenomenon (see for
instance\cite{Nishiyama:2004ju,Matos:2008ag}).

I should mention here that there is also the possibility that the SFDM
particles were produced through a \emph{non-thermal} process in the
early universe (like the typical cases of the infaton and axion
fields\cite{Kolb:1990vq}), so that the appearance of a condensed
phase could not be prevented. Such cases, however, are beyond the
purposes of this paper.

The overall conclusion of this work is then that the formation of a
relativistic BEC is the only possibility for SFDM models if the scalar
particles were in thermal equilibrium in the early Universe.


\begin{acknowledgments}
I would like to thank Andrew Liddle and Tonatiuh Matos for very useful
comments. This work was partially supported by CONACYT
(46195,47641,56946), DINPO and PROMEP-UGTO-CA-3.
\end{acknowledgments}


\bibliography{thermalref}

\end{document}